\documentclass[10pt]{article}
\usepackage[preprint]{spconf}
\usepackage{amsmath,graphicx}
\usepackage{url}
\usepackage{times}
\title{Rethinking Procedural Audio Pre-training: Source Scaling and Objective Adaptation}

\name{
Jiajun Peng$^{1,*}$,
Fengrui Liu$^{2,*,\ddagger}$,
Xinyu Liu$^{1}$,
Feng Liu$^{3,\dagger}$, \textit{Senior Member, IEEE}
}
\address{
$^{1}$School of Data Science, University of Science and Technology of China, Hefei 230026, China\\
$^{2}$School of Computer Science and Technology, East China Normal University, Shanghai 200062, China\\
$^{3}$School of Psychology, Shanghai Jiao Tong University, Shanghai 200030, China\\
$^{*}$Equal contribution. \quad $^{\dagger}$Corresponding author.\quad $^{\ddagger}$Project lead. 
}

\toappear{Submitted to ICASSP 2027}

\begin{document}
\maketitle

\begin{abstract}
Procedural audio has emerged as a viable source for transferable audio representation learning, but its design principles remain unclear.We revisit two questions: how a procedural source should be scaled, and whether training choices developed on natural audio should transfer unchanged to procedural data.Using a controlled source, we separate scale into formula-class coverage $C$ and within-class rendering diversity $I$.Experiments with FDSL and AudioMAE show that these two forms of scale provide different benefits and depend on the learning formulation and downstream task. A matched AudioMAE study further shows that procedural audio favors low mask ratios ($10$--$25\%$), whereas AudioSet-28K favors $50$--$75\%$. Shared-codebook analysis reveals lower patch diversity and stronger temporal predictability in procedural audio. These results motivate source-aware procedural pre-training, where source scaling and learning configuration are considered jointly.Code is available at \url{https://github.com/Cross-Innovation-Lab/Formula-Bank}.
\end{abstract}

\begin{keywords}
procedural audio, synthetic audio, audio pre-training, masked autoencoders
\end{keywords}

\section{Introduction}

\begin{figure}[t]
    \centering
    \includegraphics[width=\columnwidth]{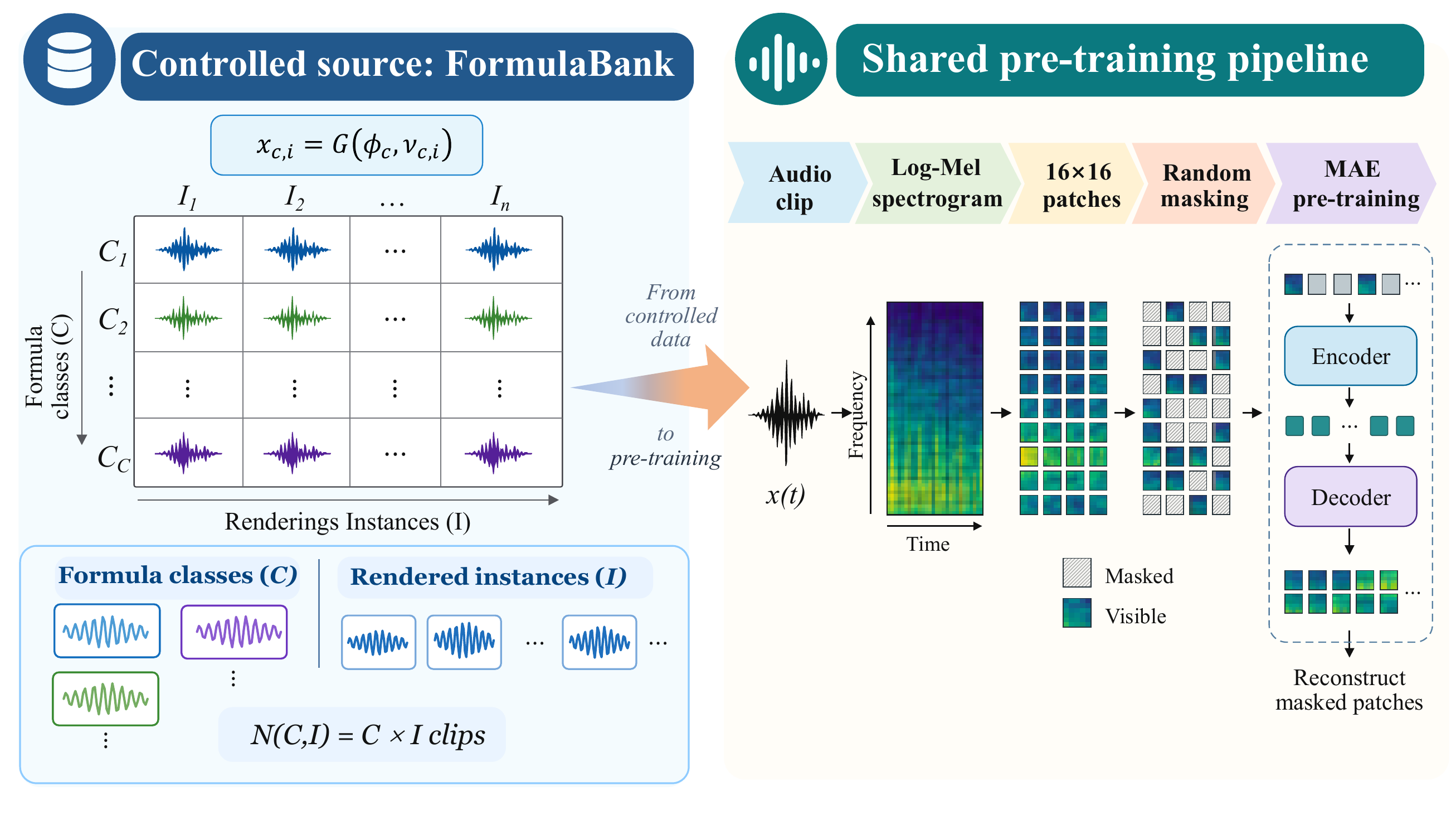}
    \caption{
    Overview of the study. 
    }
    \label{fig:overview}
\end{figure}

Formula-driven and procedural generation provide an alternative route to
representation learning without relying entirely on recorded or
human-annotated data.
FDSL showed that representations can be learned from patterns and labels
generated by mathematical formulas~\cite{kataoka2022pretraining}, while
recent work extended this idea to audio, showing that synthetic and
procedural signals can support transferable representations
~\cite{syntheticpatterns,audiopg} and complement real recordings
~\cite{prp2026}.
These studies establish the feasibility of procedural audio for
representation learning.
We therefore move from asking \emph{whether} procedural audio can work to
\emph{how procedural pre-training should be designed}.

\begin{figure*}[t]
    \centering
    \includegraphics[width=0.88\textwidth]{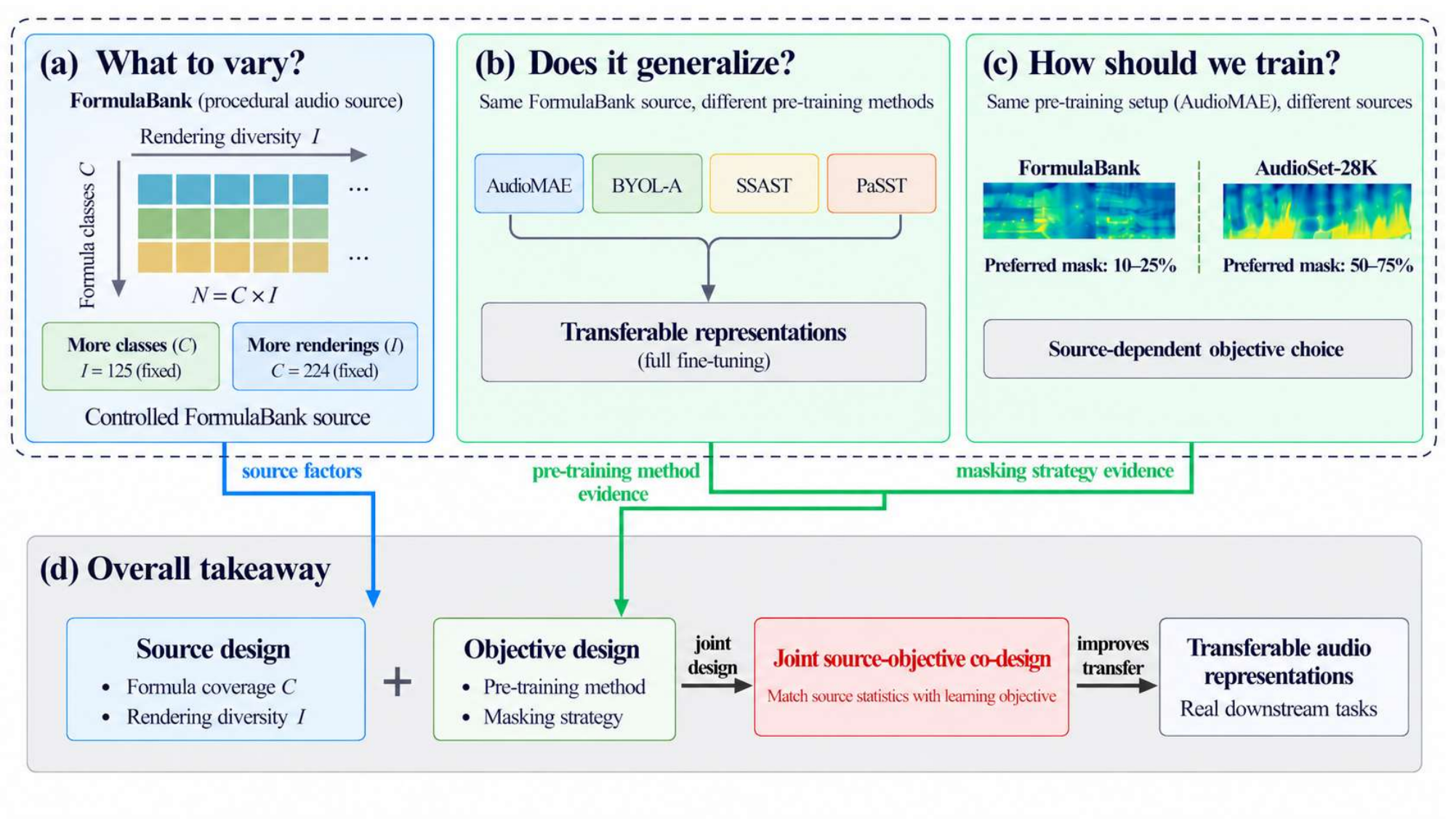}
    \caption{
    Controlled procedural source and pre-training pipeline.
    FormulaBank separates formula-class coverage $C$ from rendering
    diversity $I$, with $N(C,I)=C\times I$ clips.
    }
    \label{fig:method}
\end{figure*}

Unlike recorded datasets, procedural generation explicitly exposes how a
source grows. Increasing the number of samples can introduce new generative
structures or more acoustic realizations of existing ones, which we denote
as formula-class coverage $C$ and within-class rendering diversity $I$.
Although both increase source size, they need not provide the same learning
signal. This raises our first question: \emph{should procedural scale be
described by sample count alone, or by distinct structural and rendering
dimensions?}

The value of these dimensions may also depend on the learning formulation.
As we show~\ref{fig:overview}, FDSL and AudioMAE~\cite{audiomae} respond differently to changes
in $C$ and $I$, suggesting an interaction between source composition and the
learner. This leads to our second question: \emph{should training choices
developed on natural audio remain unchanged when the source changes?}
This is particularly relevant to masked audio pre-training, whose masking
strategies are largely developed on natural recordings
~\cite{audiomae,maskspec,baade2022maeast}.

To study these questions, we construct FormulaBank, where $C$ and $I$ can be varied independently.
We examine the two scaling axes under FDSL and AudioMAE with matched-data and matched-compute controls, and test the same source across heterogeneous pre-training paradigms~\cite{audiomae,byola,ssast,passt}.
Finally, we conduct a matched FormulaBank--AudioSet-28K~\cite{audioset} study that holds the model, data size, optimization, mask-ratio grid, and downstream evaluation fixed while changing only the pre-training source.
The results show that procedural scale is multidimensional and learner-dependent, while the preferred masking regime is strongly source-dependent.

Our main contributions are:
{\small
\begin{itemize}
    \item We disentangle procedural source scale into formula-class coverage
    and within-class rendering diversity, and show that their relative benefits
    depend on the learning formulation and downstream task.

    \item We show that the same procedural source supports heterogeneous
    audio pre-training paradigms, indicating that its utility is not tied
    to a single learning objective.

    \item A matched AudioMAE study reveals substantially different preferred
    masking regimes for procedural audio and AudioSet-28K; shared-codebook
    statistics further show lower patch diversity and stronger temporal
    predictability in procedural audio.
\end{itemize}
}

\section{methodology}

\subsection{Controlled Procedural Source}

We construct FormulaBank from the AudioPG generator~\cite{audiopg} to
separate two forms of procedural source diversity.~\ref{fig:method}
Let
\begin{equation}
    x_{c,i}=G(\phi_c,\nu_{c,i}),
\end{equation}
where $\phi_c$ denotes the generative structure of formula class $c$,
and $\nu_{c,i}$ denotes the rendering variables of its $i$-th instance.
Samples within the same class therefore share the same underlying
structure while differing in acoustic realization.

With $C$ formula classes and $I$ renderings per class, the source contains
\begin{equation}
    N(C,I)=CI
\end{equation}
clips. We use $C$ to represent \emph{structural coverage} and $I$ to
represent \emph{within-structure rendering diversity}. This construction
allows the two forms of procedural scale to be varied independently.
\begin{table*}[t]
\centering
\caption{Scaling formula-class coverage and rendering diversity under FDSL
and AudioMAE pre-training. Results within each pre-training setting use the
same downstream evaluation protocol.}
\label{tab:scaling}
\setlength{\tabcolsep}{3.2pt}
\resizebox{\textwidth}{!}{
\begin{tabular}{cc|ccccc|ccccc}
\hline
& &
\multicolumn{5}{c|}{FDSL} &
\multicolumn{5}{c}{AudioMAE} \\
Axis & Scale &
ESC-50 Acc. $\uparrow$ & US8K Acc. $\uparrow$ & FSD50K mAP $\uparrow$ &
SCv2 Acc. $\uparrow$ & AS-20K mAP $\uparrow$ &
ESC-50 Acc. $\uparrow$ & US8K Acc. $\uparrow$ & FSD50K mAP $\uparrow$ &
SCv2 Acc. $\uparrow$ & AS-20K mAP $\uparrow$ \\
\hline

\multicolumn{12}{l}{\textit{Formula-class coverage ($I=125$)}} \\
& $C=14$  & 76.50 & 76.11 & 0.4507 & 97.30 & 0.1330
            & 71.75 & 78.61 & 0.4431 & 96.79 & 0.1163 \\
& $C=28$  & 74.50 & 77.54 & 0.4599 & 97.08 & 0.1382
            & 76.25 & 81.12 & 0.4854 & 96.66 & 0.1579 \\
& $C=56$  & 76.00 & 75.39 & 0.4720 & \textbf{97.42} & 0.1310
            & \textbf{78.50} & \textbf{84.59} & 0.4981 & \textbf{97.21} & 0.1657 \\
& $C=112$ & 76.25 & 80.76 & \textbf{0.4957} & 97.23 & 0.1366
            & 77.50 & 83.75 & 0.4921 & 97.19 & 0.1746 \\
& $C=224$ & \textbf{77.00} & \textbf{82.92} & 0.4919 & 97.34 & \textbf{0.1423}
            & 77.50 & 83.99 & \textbf{0.5073} & 97.16 & \textbf{0.1895} \\
\hline

\multicolumn{12}{l}{\textit{Rendering diversity ($C=224$)}} \\
& $I=16$   & 76.75 & 76.22 & 0.4742 & \textbf{97.40} & 0.1329
             & 66.50 & 75.51 & 0.4365 & 96.52 & 0.1238 \\
& $I=32$   & 76.25 & 81.36 & 0.4785 & 97.22 & 0.1373
             & 72.75 & 80.05 & 0.4732 & 97.04 & 0.1549 \\
& $I=64$   & 75.50 & 81.00 & 0.4805 & 97.06 & 0.1424
             & 77.75 & 84.11 & 0.4901 & 97.22 & 0.1680 \\
& $I=125$  & 77.00 & \textbf{82.92} & 0.4919 & 97.34 & 0.1423
             & 77.50 & 83.99 & 0.5073 & 97.16 & 0.1895 \\
& $I=250$  & 77.25 & 81.72 & 0.4865 & 97.18 & 0.1506
             & 82.25 & 83.99 & 0.5221 & \textbf{97.42} & 0.1966 \\
& $I=500$  & \textbf{78.50} & 80.76 & \textbf{0.4928} & 97.27 & \textbf{0.1535}
             & 81.25 & \textbf{85.90} & 0.5303 & 97.27 & \textbf{0.2017} \\
& $I=1000$ & 77.75 & 79.69 & 0.4868 & 97.07 & 0.1480
             & \textbf{82.75} & \textbf{85.90} & \textbf{0.5413} & 97.23 & 0.2006 \\
\hline
\end{tabular}}
\end{table*}

\subsection{Scaling Procedural Sources}

We study the two scaling axes separately:
\begin{align}
    I &= 125,\quad
    C\in\{14,28,56,112,224\},\\
    C &= 224,\quad
    I\in\{16,32,64,125,250,500,1000\}.
\end{align}

The same source configurations are evaluated under FDSL~\cite{kataoka2022pretraining}
and AudioMAE~\cite{audiomae}, allowing us to examine whether the value of
structural coverage and rendering diversity depends on the learning formulation.
We additionally use fixed-data and fixed-compute controls to separate changes
in source composition from additional training exposure.

\subsection{Generality and Source Adaptation}

We first test whether the procedural source transfers across different
representation-learning paradigms by pre-training AudioMAE~\cite{audiomae},
BYOL-A~\cite{byola}, SSAST~\cite{ssast}, and PaSST~\cite{passt} on the same
FormulaBank source.

We then isolate the effect of source domain using a matched AudioMAE setup.
FormulaBank and a 28K-clip AudioSet subset~\cite{audioset} use the same model,
optimization, input representation, and downstream evaluation, while only the
pre-training source is changed. For both sources, we sweep
\begin{equation}
    r\in\{.10,.25,.50,.75,.80,.90\},
\end{equation}
where $r$ is the fraction of masked spectrogram patches.

\subsection{Source Statistics}

To relate masking behavior to source statistics, we quantize patches from
FormulaBank and AudioSet-28K using a shared codebook.
Let $Z$ denote the resulting discrete patch code. We measure patch diversity by
\begin{equation}
    H(Z)=-\sum_k p(k)\log_2 p(k),
\end{equation}
and local temporal predictability by
\begin{equation}
    H_t = H(Z_{t+1,f}\mid Z_{t,f}).
\end{equation}
Lower $H_t$ indicates stronger predictability between adjacent time patches.
\section{Experiments}

\subsection{Experimental Setup}
\begin{figure*}[t]
    \centering
    \includegraphics[width=.28\textwidth]
    {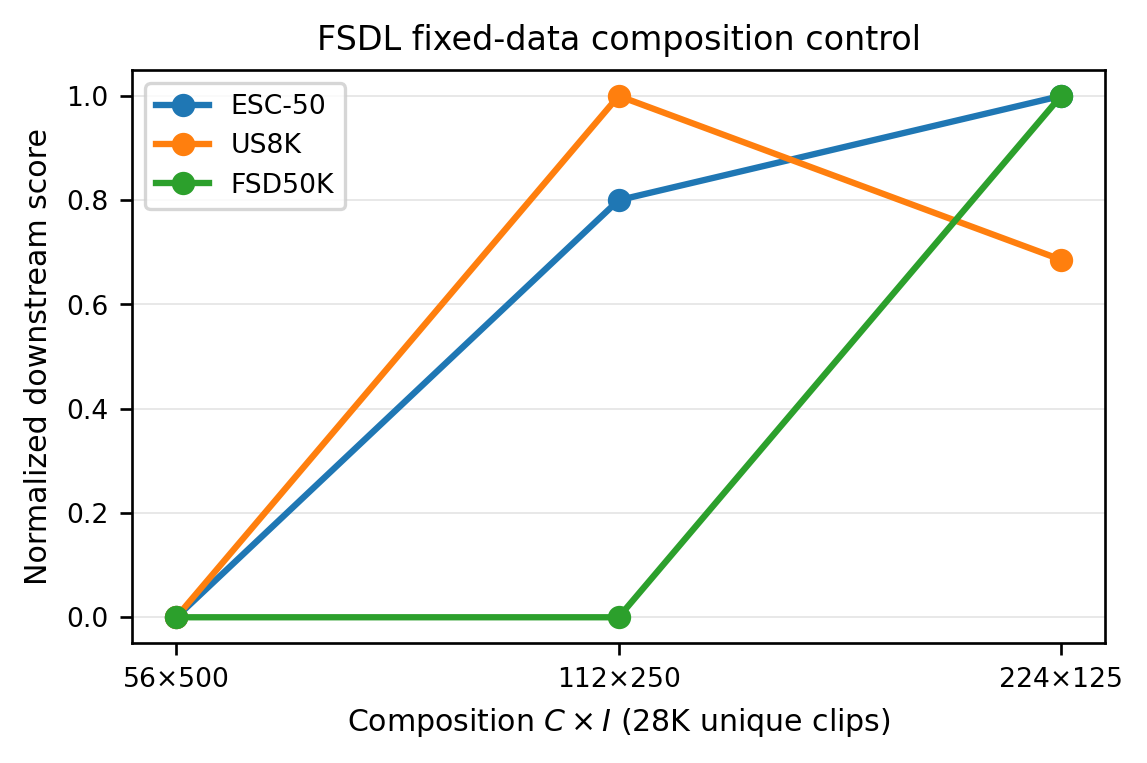}
    \hfill
    \includegraphics[width=.28\textwidth]
    {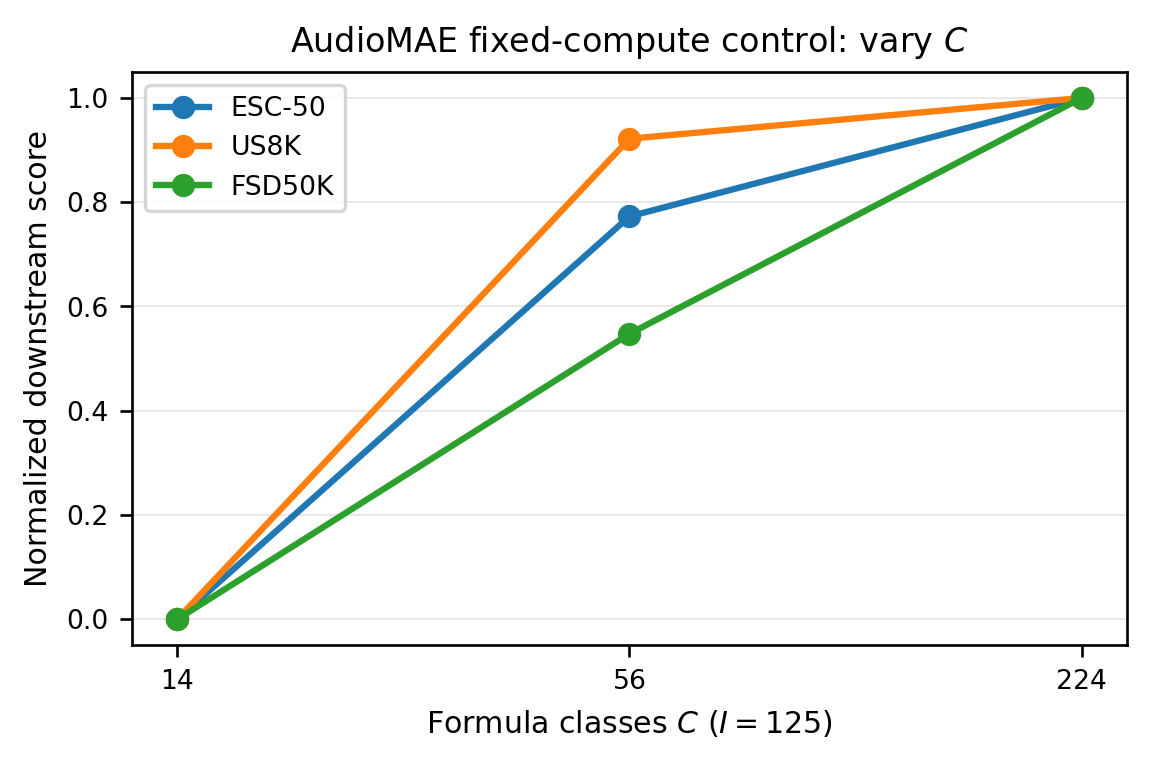}
    \hfill
    \includegraphics[width=.28\textwidth]
    {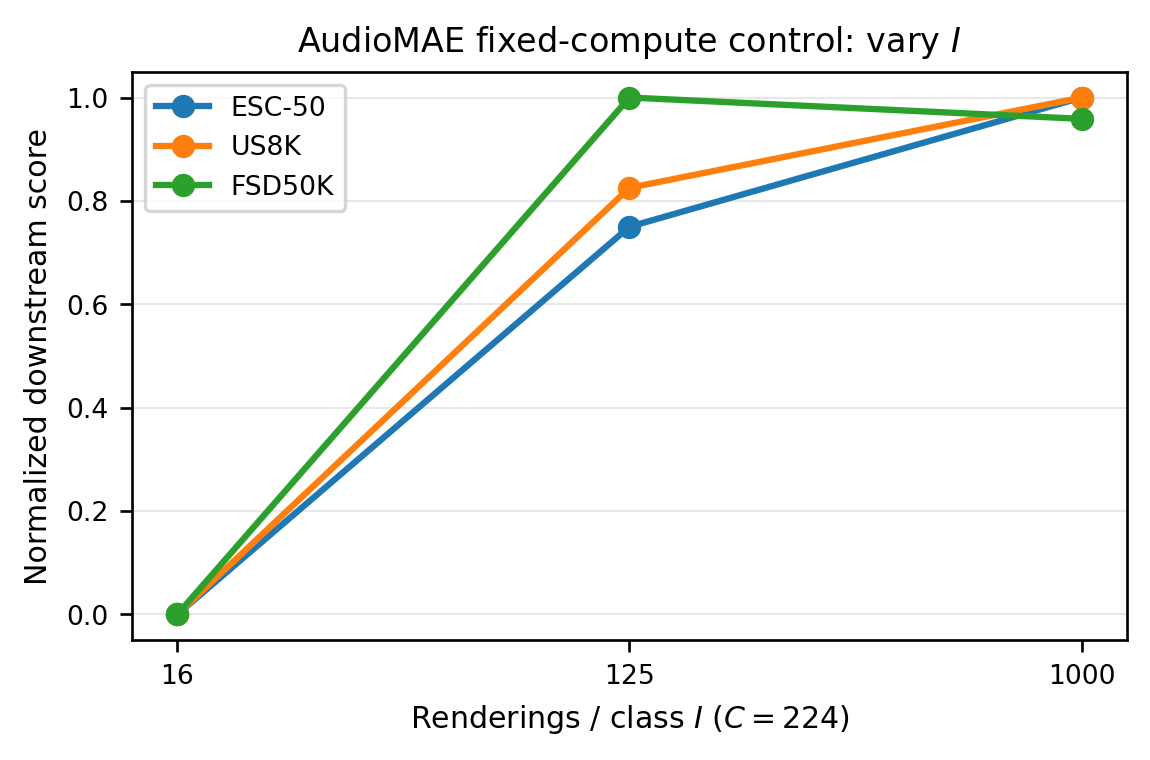}
    \caption{Controlled-budget scaling trends.
    Left: FDSL with 28K unique clips, varying source composition from
    $56\times500$ to $224\times125$.
    Middle: AudioMAE with 7M clip presentations, increasing formula-class
    coverage at $I=125$.
    Right: AudioMAE under the same compute budget, increasing rendering
    diversity at $C=224$.
    Scores are normalized within each downstream task.}
    \label{fig:scaling_control}
\end{figure*}
\textbf{Datasets and metrics.}
Pre-training uses FormulaBank as the procedural source and AudioSet-28K
~\cite{audioset} as the matched natural-audio control where applicable.
Downstream evaluation is conducted on ESC-50~\cite{esc50},
UrbanSound8K (US8K)~\cite{us8k}, FSD50K~\cite{fsd50k},
Speech Commands v2 (SCv2), and AudioSet-20K (AS-20K).
We report classification accuracy for ESC-50, US8K, and SCv2,
and mean average precision (mAP) for FSD50K and AS-20K.

\textbf{Implementation details.}
AudioMAE experiments use 128-bin log-Mel spectrograms of size
$1024\times128$ with non-overlapping $16\times16$ time--frequency patches.
FormulaBank scaling and masking experiments use ViT-S, while the
cross-method study retains each method's native encoder and training recipe.
All downstream models are fully fine-tuned using the same protocol for a
given downstream dataset.
\begin{figure*}[t]
    \centering
    \includegraphics[width=.8\textwidth]{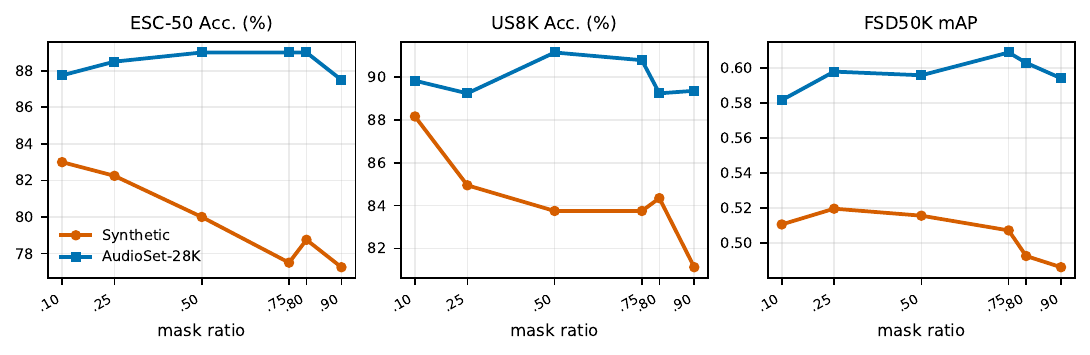}
    \caption{Source-dependent masking under full fine-tuning.
    FormulaBank favors low masking ($.10$--$.25$), whereas the matched
    AudioSet-28K control favors moderate-to-high masking ($.50$--$.75$).
    The AudioMAE architecture, clip count, mask-ratio grid, and
    downstream evaluation are held fixed across the two sources.}
    \label{fig:masking}
\end{figure*}

\subsection{Multidimensional Procedural Scaling}

We first examine how procedural pre-training scales along two controllable
source dimensions: formula-class coverage $C$ and within-class rendering
diversity $I$. We fix $I=125$ when varying $C$, and fix $C=224$ when varying
$I$, under both FDSL~\cite{kataoka2022pretraining} and
AudioMAE~\cite{audiomae}.

Table~\ref{tab:scaling} shows that the two forms of scale are not
equivalent. Under FDSL, broader formula-class coverage produces the
clearer trend, whereas rendering diversity yields smaller and less
consistent changes. AudioMAE shows a different response: coverage gives
rapid early gains followed by task-dependent saturation, while rendering
diversity provides broader improvements across downstream tasks.

To separate source composition from additional training exposure, we use
matched-budget controls: FDSL fixes the source size to 28K clips, while
AudioMAE fixes optimization exposure to 7M clip presentations.

Figure~\ref{fig:scaling_control} shows that these trends persist under
controlled budgets, supporting a multidimensional view of procedural scale
whose axes depend on both the learning formulation and downstream task.

\subsection{Generality Across Learning Paradigms}

We next test whether the procedural source is tied to a particular
pre-training objective. The same FormulaBank source is used with
AudioMAE~\cite{audiomae}, BYOL-A~\cite{byola},
SSAST~\cite{ssast}, and PaSST~\cite{passt}, while retaining each
method's native encoder and training recipe.

\begin{table}[t]
\centering
\caption{Cross-method transfer from FormulaBank pre-training.
Each method uses its native configuration and is evaluated by full
fine-tuning.}
\label{tab:crossmethod}
\setlength{\tabcolsep}{2.8pt}
\resizebox{\columnwidth}{!}{
\begin{tabular}{lcccccc}
\hline
Method & Encoder & ESC-50 & US8K & FSD50K & SCv2 & AS-20K \\
\hline
AudioMAE & ViT-B
& 86.25 & 85.78 & .5649 & 97.77 & .241169 \\
BYOL-A & AudioNTT2022-CNN
& 75.00 & 76.82 & .4660 & 96.70 & .152646 \\
SSAST & ViT-B
& 81.00 & 81.36 & .5167 & 96.72 & .199608 \\
PaSST-noImageNet & ViT-B
& 75.50 & 77.66 & .5156 & 96.81 & .179001 \\
Scratch & ViT-B
& 69.50 & 73.36 & .4090 & 96.74 & .136188 \\
\hline
\end{tabular}}
\end{table}

Table~\ref{tab:crossmethod} shows that all four learning paradigms
produce transferable representations from the same procedural source.
Since the architectures and native training recipes differ, this
experiment evaluates source compatibility rather than ranking the
pre-training methods.

\subsection{Source-Dependent Masking and Source Statistics}

We finally ask whether the learning configuration should adapt when the
pre-training source changes. FormulaBank and a matched 28K-clip
AudioSet subset~\cite{audioset} are compared under the same AudioMAE
setup, with architecture, optimization, mask-ratio grid, and downstream
evaluation held fixed.

Figure~\ref{fig:masking} shows a clear source-dependent shift.
FormulaBank performs best at low mask ratios ($r=.10$--$.25$), whereas
AudioSet-28K favors $r=.50$--$.75$. This trend is consistent across
downstream tasks and repeated runs.
\begin{table}[t]
\centering
\caption{Shared-codebook patch entropy in bits. Higher values indicate
greater patch diversity or lower short-range predictability.}
\label{tab:entropy}
\setlength{\tabcolsep}{4pt}
\begin{tabular}{lccc}
\hline
Source & Marginal $H$ & Within-clip $H$ & Temporal $H_t$ \\
\hline
FormulaBank & 3.140 & 2.104 & .715 \\
AudioSet-28K & 6.701 & 4.941 & 3.807 \\
\hline
\end{tabular}
\end{table}
To characterize this difference, we compare the two sources in a shared
discrete patch space.Table~\ref{tab:entropy} shows that FormulaBank has lower patch diversity
and substantially stronger local temporal predictability than AudioSet-28K.
This suggests that the same masking ratio induces different prediction
difficulty across source domains, motivating source-dependent masking.
\section{Conclusion}
Overall, these results motivate source--objective co-design: masking should be adapted to the statistics of the pre-training source rather than inherited as a source-independent default.

\bibliographystyle{IEEEbib}
\bibliography{refs}
\end{document}